\documentclass{article}

        \usepackage[nonatbib,final]{neurips_2020}

\usepackage[utf8]{inputenc} 
\usepackage[pdftex]{graphicx}
\usepackage[T1]{fontenc}    
\usepackage{hyperref}       
\usepackage{url}            
\usepackage{booktabs}       
\usepackage{amsfonts}       
\usepackage{nicefrac}       
\usepackage{microtype}      
\usepackage{mathtools}
\usepackage[square,numbers]{natbib}

\title{Binary Segmentation of Seismic Facies Using Encoder-Decoder Neural Networks}

\author{%
    Gefersom Lima\thanks{Graduate Program in Applied Computing.}~, Gabriel Ramos$^*$, Sandro Rigo$^*$, Felipe Zeiser$^*$, Ariane da Silveira\thanks{Graduate Program in Geology.} \\
    Universidade do Vale do Rio dos Sinos, Brazil\\
    \{gefersom@edu., gdoramos@, rigo@, felipezeiser@edu., ariane@\}unisinos.br
}

\begin{document}

\maketitle


\begin{abstract}
The interpretation of seismic data is vital for characterizing sediments' shape in areas of geological study. In seismic interpretation, deep learning becomes useful for reducing the dependence on handcrafted facies segmentation geometry and the time required to study geological areas. This work presents a Deep Neural Network for Facies Segmentation (DNFS) to obtain state-of-the-art results for seismic facies segmentation. DNFS is trained using a combination of cross-entropy and Jaccard loss functions. 
Our results show that DNFS obtains highly detailed predictions for seismic facies segmentation using fewer parameters than StNet and U-Net.
\end{abstract}

\section{Introduction}

A crucial task in hydrocarbon exploration refers to seismic interpretation, which aims at investigating the organization of the soil's rocky layers and identifying regions with structures capable of storing petroleum and gas~\cite{Ashcroft2011}. Seismic data is manually interpreted in conventional workflows by human experts (who mark the transitions between seismic wave reflection patterns in a time-consuming way) \cite{LiuJervisLiNivlet2020}, or it is made with auto-tracking tools that identify patterns in the seismic data. A significant drawback of these tools is that they cannot enhance their results without an interpreter's direct interference. To mitigate this problem, we made use of neural networks for segmenting seismic facies geometries.

The DNFS (Deep Neural Network for Facies Segmentation) proposed in this work is a convolution neural network based on an encoder-decoder structure. This network can learn essential characteristics that enable constructing an output image with segmented seismic facies separated by black lines in white background (Figure~\ref{fig:data_dnfs}). DNFS builds upon U-Net \cite{RonnebergerFischerBrox2015} and StNet \cite{HaibinDengliangGhassan2019} network properties. We created a dataset on these black lines (there is no dataset widely used for binary segmentation of seismic facies) and a composite loss function formed by the combination of cross-entropy and Jaccard loss for training. This function aims to consider the pixels' spatial class distribution in the result predictions, which is not attainable using either cross-entropy or Jaccard alone. 

To validate DNFS, we performed an extensive experimental evaluation to tune the hyperparameters and optimize the loss function's coefficient. Also, we created a specific dataset for segmenting seismic facies based on binary segmentation. Our results show that DNFS is capable of segmenting an arbitrary number of seismic facies because it only focuses on the transitions between them.


\section{Related Work}

An important task within seismic interpretation refers to the analysis of seismic facies~\cite{DumayFournier1988}. For assisting this task with deep learning, researchers have been applying CNNs for seismic interpretation \cite{YunzhiXinmingSergey2019, WuLiangShiGengFomel2019, LiZhouWangWu2020}. Most works focus on a single seismic event, such as salt dome \cite{WaldelandJensenGeliusSolberg2018,  Muhammad2020, ZengJiangChen2019}, and fault detection \cite{HuangDongClee2017, XiongJiMaWangBenHassanAliLuo2018, PochetDinizLopesGattass2019}. Other works like \cite{Li2018, AlaudahMichalowicz2019, HaibinDengliangGhassan2019} used CNNs in an encoder-decoder model for automated detection and classification of multi-class geologic structure elements. \citet{HaibinDengliangGhassan2019} presented an encoder-decoder neural network (called StNet) for segmenting twelve seismic facies geometries. 

When different seismic facies need to be segmented simultaneously, segmenting only one seismic facies at a time is not usually the best choice because the predicted results could not be easily merged. To avoid trying to target different seismic regions using different classes, even when these locations are the same, we could train neural networks to focus only on the separating region between adjacent seismic facies. Thus, the training would be based on binary segmentation, decreasing the number of neural network parameters.


\section{Method and Results}

Our problem involves identifying seismic facies in seismic data. We thus created a dataset where the transitions between seismic facies were exposed through black lines. This characteristic made our problem a binary-segmentation one. Based on this aspect, we created a neural network (DNFS) and trained it with the composite loss function shown below, where $\psi$ weighs the importance of cross-entropy and Jaccard loss in the linear composite function.
\begin{equation}
    Loss = \psi * CrossEntropy + (1 -  \psi) * Jaccard
\label{eq:composit_loss_function}
\end{equation}

As evaluation metrics, we used intersection over union (IoU) and the percentage of correct black pixel predictions. The architecture of DNFS is composed of features of StNet and U-Net neural networks. We applied the same smooth transition of feature maps of U-Net and the transpose convolution layers of StNet on DNFS (on the decoder part). To connect the encoder and decoder parts, instead of using two layers as is done on StNet and U-Net, only one layer is used in DNFS, which helped decrease the total number of network parameters.

\begin{figure}[b]
	\centering
	\includegraphics[width=1.0\linewidth]{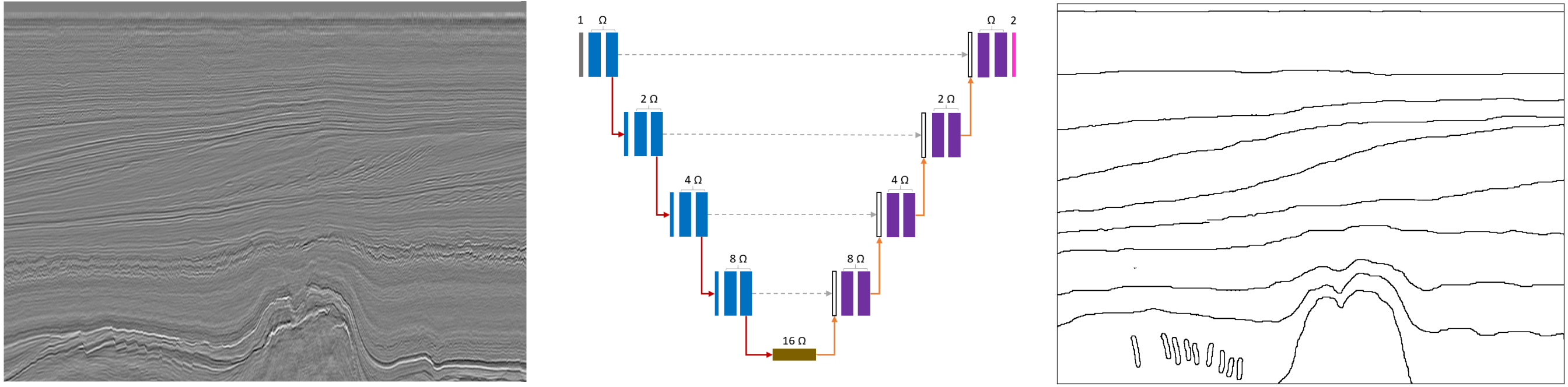}
	\caption{On the left, an example of seismic input image. In the center, the propose DNFS architecture. On the right, a prediction made by DNFS.}
	\label{fig:data_dnfs}
\end{figure}

Figure~\ref{fig:data_dnfs} shows the resulting DNFS architecture. This combination resulted from eighteen experiments with different neural network configurations (six variations of DNFS, StNet, and U-Net). Each of them was constructed by scaling the number of filters in the layers; the scale factors were: 4, 8, 16, 32, 64, and 128. We considered the total number of parameters, time training, and the percentage of correct black pixels for choosing the best neural network. 

After training these neural networks with train, validate, and test subsets of the created dataset, we noted that the total number of parameters varied from 24.917 to 124.164.36, the time training from five minutes to nine hours, and the percentage of correct black pixels, from 0 to 87\%. We chose the DNFS with multiplier eight among the other networks because it can be used with environments where there is little memory (it only consumes 3MB) and processing power (it has 341.001 total of parameters). Moreover, even with DNFS having fewer parameters than StNet (1.505.69) and U-Net (31.059.085), it could obtain  85.89\% of correctly predicted black pixels while StNet obtained 83.3\%, and U-Net, 86.67\%.

Based on current results, we hope to provide a neural network capable of being trained in a short time to be used in the interpretation workflow, reducing the time and effort required to interpret seismic facies. For enhancing our results, we could use 1x1 convolution on the skip connections and modify the Equation \eqref{eq:composit_loss_function} to further penalize the incorrect prediction of black pixels.

\bibliographystyle{plainnat}
\bibliography{cas-refs}

\end{document}